\documentclass[12pt,preprint]{aastex}
\usepackage{graphicx}
\begin{document}

\shortauthors{Prieto et~al.}

\title{The Rise of the Remarkable Type~IIn Supernova SN~2009ip}

\author{Jos\'e~L.~Prieto\altaffilmark{1}, J.~Brimacombe\altaffilmark{2,3}, A.~J.~Drake\altaffilmark{4},
S.~Howerton\altaffilmark{5}}

\altaffiltext{1}{Department of Astrophysical Sciences, Princeton University, NJ 08544, USA} 
\altaffiltext{2}{James Cook University, Cairns, Australia}
\altaffiltext{3}{Coral Towers Observatory, Unit 38 Coral Towers, 255 Esplanade, Cairns 4870, Australia}
\altaffiltext{4}{California Institute of Technology, 1200 E. California Blvd., CA 91225, USA}
\altaffiltext{5}{1401 South A, Arkansas City, KS 67005, USA}

\begin{abstract}
  Recent observations by Mauerhan et al. have shown the unprecedented
  transition of the previously identified luminous blue variable (LBV) and
  supernova impostor SN~2009ip to a real Type~IIn supernova (SN)
  explosion. We present high-cadence optical imaging of SN~2009ip
  obtained between 2012 UT Sep.~23.6 and Oct.~9.6, using $0.3-0.4$~meter
  aperture telescopes from the Coral Towers Observatory in Cairns,
  Australia. The light curves show well-defined phases, including very
  rapid brightening early on (0.5~mag in 6~hr observed during the night
  of Sep.~24), a transition to a much slower rise between Sep.~25 and
  Sep.~28, and a plateau/peak around Oct.~7. These changes are
  coincident with the reported spectroscopic changes that, most likely,
  mark the start of a strong interaction between the fast SN ejecta and
  a dense circumstellar medium formed during the LBV eruptions observed
  in recent years. In the 16-day observing period SN~2009ip brightened
  by 3.7~mag from $I=17.4$~mag on Sep.~23.6 ($M_I \simeq -14.2$) to
  $I=13.7$~mag ($M_I \simeq -17.9$) on Oct.~9.6, radiating $\sim 2\times
  10^{49}$~erg in the optical wavelength range. Currently, SN~2009ip is
  more luminous than most Type~IIP SNe and comparable to other Type~IIn
  SNe. 
\end{abstract}


\section{Introduction}

Establishing an observational mapping between the properties of
core-collapse supernovae (ccSNe) explosions (and related luminous
outbursts) and local populations of massive stars is key for
constraining stellar evolution theory (e.g., Langer 2012). Arguably, the
best link so far comes from the direct detections of red-supergiants
with main-sequence masses $\rm M\simeq 8-15$~M$_{\odot}$ as progenitors
of Type~IIP supernovae (e.g., Smartt 2009, and references therein), the
most common kind of ccSNe. However, direct detections of SN progenitors
with higher masses ($M\gtrsim 20$~M$_{\odot}$) has been elusive (e.g.,
Kochanek et al. 2008; Smartt 2009). 

The detection of a very luminous star ($M_V\sim -10$~mag) identified in
pre-explosion images at the location of the Type~IIn\footnote{Dominated
by narrow H lines in emission, which indicates a strong interaction
between the fast ejecta and a dense circumstellar medium (CSM).}
SN~2005gl provided the first direct evidence for a very massive H-rich
star ($\gtrsim 50$~M$_{\odot}$) that exploded as a luminous ccSN (Gal-Yam
et al. 2007; Gal-Yam \& Leonard 2009). These observations lend support
for the connection made initially by Smith et al. (2007) between one of
the most luminous Type~IIn SN 2006gy and very massive ($\sim
100$~M$_{\odot}$) luminous blue variable (LBV) stars like
$\eta$~Carinae. There is now mounting evidence of this
connection (e.g., SN~2006jc, Pastorello et al. 2007; SN~2010jl, Smith et al. 2011a; SN~1961V, Kochanek et al. 2011, 
Smith et al. 2011b), which challenges massive stellar evolution theory in at
least two important ways: 1) the most massive stars are not expected to
be H-rich at the time when they explode as ccSNe, at least with normal
mass-loss prescriptions; and 2) these stars have to loose up to a few
solar masses of H-rich material in eruptions just before (months to
decades) the core-collapse explosion in order to make them visible as
Type~IIn SNe. On this last point of timing strong eruptive mass-loss
with the core-collapse explosion, there is some very interesting
theoretical work, but its still very early days (e.g., Woosley et al. 2007; 
Arnett \& Meakin 2011; Quataert \& Shiode 2012; Chevalier 2012).

Luckily, in the last month nature has been kind and has provided the
best case to date of a very massive star exploding as a bright
core-collapse supernova: SN~2009ip. This transient was discovered on
2009 Aug.~29 by the CHASE survey in the outskirts of the galaxy NGC~7259
(Maza et al. 2009) and was initially given a SN name by the IAU.
However, detailed photometric and spectroscopic studies by Smith et al.
(2010) and Foley et al. (2011) showed that it never quite reached
supernova status. It was actually a supernova impostor (e.g., Humphreys
\& Davidson 1994, Kochanek et al. 2012; with peak absolute mag. $\sim
-14$, similar to $\eta$~Carinae's Great Eruption, but significantly
shorter timescale), in this case a massive LBV with estimated mass $\rm
M\simeq 50-80$~M$_{\odot}$ (constrained from pre-discovery imaging) that
had an eruptive mass-loss event. The Catalina Real-Time Transient
Survey (CRTS) discovered another outburst of comparable amplitude during
2010 (Drake et al. 2010) and a third outburst on 2012 July 24 (Drake et
al. 2012), the brightest so far. The initial optical spectra obtained by
Foley et al. (2012) on Aug.~26 showed narrow emission lines of H and
He~I, consistent with the spectra obtained in the 2009 outburst.
However, an unprecedented change of the spectrum was reported by Smith
\& Mauerhan (2012a). On Sep.~15 and 16 they had detected very broad
emission lines with P-Cygni profiles, consistent with features observed
in real Type~II SNe; the massive LBV star had probably exploded as a supernova in
real time! Another dramatic change in the spectral properties was
reported by Smith \& Mauerhan (2012b), by Sep.~28 the broad P-Cygni
features had mostly disappeared, leaving behind narrow features
characteristic of Type~IIn SNe. All the details of their discovery and their 
follow-up observations can be found in Mauerhan et al. (2012b; hereafter M12b).
Pastorello et al. (2012; hereafter P12) has also presented detailed 
follow-up observations of SN~2009ip and propose that the latest outburst
might not be the final core-collapse explosion. 

In this paper we present high-cadence optical observations of SN~2009ip
that clearly resolve the brightening initially reported by Brimacombe
(2012). In Section~\S\ref{sec1} we discuss the observations, data
reduction, and extraction of optical light curves using difference
imaging photometry. In Section~\S\ref{sec2} we present an analysis of
the different phases of the light curve. In Section~\S\ref{sec3} we
present a discussion of the observed properties. In Section~\S\ref{sec4}
we present the conclusions of this work. We adopt a distance of 20.4~Mpc
($\mu=31.55$~mag) and Galactic extinction of $A_V = 0.06$~mag towards
SN~2009ip (M12b) throughout the paper. All the dates presented in this
paper are UT.  

\section{Observations and Photometry}
\label{sec1}

Imaging of the field of SN~2009ip was obtained by one of us (J.~B.)
between 2012 Sep.~23.6 and Oct.~9.6 from the Coral Towers Observatory
(Cairns, Australia). The data were collected from two different
telescopes using different broad-band filters: 33-cm RCOS (with an $R$
filter) and 41-cm RCOS (with an IR luminance filter, which has sharp blue
cutoff at $700$~nm). The CCD cameras used in both telescopes are
identical 3k$\times$2k SBIG STL6K, which give a total field of view of
$25\arcmin\times 17\arcmin$ (1\arcsec/pix platescale binned $2\times2$).
All the images were obtained using an exposure time of 900~sec.
A section of an IR image of SN~2009ip obtained on Sep.~24 is shown in
Figure~1.

We used the software MaxIm DL (Version 4.62) to complete the initial
data reduction, which consists of bias subtraction, dark subtraction,
and flat-fielding. After visually inspecting all images and rejecting
frames affected by bad tracking, we kept 85 ($R$) and 118 (IR) frames
for further photometric analysis. We used the image subtraction package
ISIS2 (Alard 2000) to extract the fluxes of SN~2009ip in individual
frames following the steps described in Hartman et al. (2004). As
reference images, we used stacks of $5-10$ images with good seeing
obtained on Sep.~24-25. We carried out PSF photometry with the DAOPHOT
II package (Stetson 1992) to subtract the SN flux and used the resulting
SN-subtracted images as the final reference frames.

The SN fluxes obtained with ISIS2 in the $R$ and IR images were
calibrated to standard $R$ and $I$ magnitudes, respectively, using data
from the AAVSO Photometric All-Sky Survey (APASS\footnote{\tt
http://www.aavso.org/apass}). We obtained from APASS calibrated Sloan r'
and i' magnitudes of 5 isolated stars in the field of SN~2009ip. 
We converted these magnitudes in the SDSS photometric system
to standard $RI$ (Johnson/Kron-Cousins) photometry using the
transformation equations obtained by R.~Lupton\footnote{\tt
http://www.sdss.org/dr7/algorithms/sdssUBVRITransform.html\#Lupton2005}.
In order to estimate the zeropoints, we performed aperture photometry of
the 5 local standard stars in the reference images using a 9\arcsec
radius aperture with the IRAF task {\tt phot}. The resulting average
zeropoints in the $RI$ filters have a standard deviation of 0.02~mag.
The final photometry of SN~2009ip calibrated using these zeropoints is
presented in Table~1 and the light curves are shown in Figure~1. The
errors in each magnitude estimate include poisson errors from ISIS2 as
well as the estimated standard error in the photometric zeropoints. 
Our photometry is in excellent agreement with the results presented 
in P12, with mean differences during the period of observations of 
$-0.01 \pm 0.01$~mag in $R$ and $-0.02 \pm 0.03$~mag in $I$. 

\section{Analysis}
\label{sec2}

The high cadence and number of observations allow us to clearly resolve
the brightening of SN~2009ip, initially reported in Brimacombe (2012) 
and Prieto et al. (2012). The optical light curves presented in Figure~2 show 
well-defined phases: 1) approximately constant magnitude (Sep.~23); 2) very 
rapid brightening in a period of hours (Sep.~$24-25$);  3) turn-over to 
significantly slower brightening (Sep.~$25-28$); and 4) slow brightening, 
reaching peak magnitude (Sep.~30 $-$ Oct.~9). We describe these phases in 
more detail below.  

Between Sep.~23.56 and 23.66, the magnitude of SN~2009ip is consistent
with being constant at $I=17.39 \pm 0.05$~mag ($M_I \simeq -14.2$).
However, between Sep.~23.66 and 24.45 SN~2009ip brightened by $\Delta I \simeq
0.65$~mag and continued a very rapid brightening throughout the 6~hrs of
continuous imaging obtained during the night of Sep. 24. In this period,
it brightened between $I=16.7$ (Sep.~24.45) and $I=16.2$~mag
(Sep.~24.70). A linear fit to the $I$-band rise on Sep.~24 gives a slope
of $2.30\pm 0.10$~mag~day$^{-1}$.

Between Sep.~24.70 and 25.38 the SN brightened by $\Delta I \simeq
1.0$~mag, reaching $I=15.1$~mag ($M_I \simeq -16.5$). The light curve 
turned over to a slower brightening on Sep.~25. A linear fit to the 
data obtained during the night of Sep.~25 gives a slope of $0.79 \pm
0.05$~mag~day$^{-1}$. The turn-over can be seen clearly in both the $R$ and 
$I$-band photometric data obtained between Sep.~25 and Sep.~28 (see
Figure~2), reaching $I=14.1$~mag ($M_I \simeq -17.5$) by Sep.~28.47.

The latest part of the light curve between Sep.~30 and Oct.~7 shows a
slow and steady brightening at a rate of $0.037\pm 0.001$~mag~day$^{-1}$
($\rm rms = 0.01$~mag). However, in our most recent images obtained on
Oct.~9.6 the SN faded by $\sim 0.03$~mag with respect to the images
obtained on Oct.~7.6. The magnitude of SN~2009ip on Oct.~9.6 was
$I=13.70 \pm 0.02$~mag ($M_I \simeq -17.9$). The $R-I$ color is
consistent with being constant ($R-I= -0.02 \pm 0.03$) during the period
between Sep.~25 and Oct.~9.

\section{Discussion}
\label{sec3}

Our photometric observations, obtained during a 16-day period between
Sep.~23.6 and Oct.~9.6 show that the Type~IIn SN~2009ip brightened by
3.7~mag in the optical to an absolute magnitude $M_R = M_I \simeq -17.9$
(see also Margutti et al. 2012b; M12b; P12) and it has reached peak
magnitude around Oct.~7 (see Figure~2). Its peak luminosity is
in the observed range of Type~II SNe; in fact it is more luminous than
most Type~IIP SNe (e.g., Li et al. 2011). The comparison to other
Type~IIn SNe is less straightforward since the rates are lower and the
samples are incomplete. However, there are several well-studied Type~IIn
SNe with peak absolute magnitudes around $M\sim-18$ (e.g., Kiewe et al.
2012; Stritzinger et al. 2012; Roming et al. 2012; Mauerhan et al. 2012a). 

We can estimate the total radiated energy during the period of the
observations presented here. We need an estimate of the bolometric
correction to convert the $I$-band luminosities to bolometric
luminosities and we do this in two ways. First, we use the observed
constant $R-I$ color, corrected for Galactic extinction, and fit a
black-body to estimate the luminosity ratio $L_{\rm bol}/L_I \simeq 7$
and $T_{\rm eff}\approx 14500$~K. Second, we use the longer wavelength
coverage (from near-UV to $V$-band) of the Swift observations presented
in Margutti et al. (2012b) and fit a black-body to the
extinction-corrected fluxes, obtaining $L_{\rm bol}/L_I \simeq 13$ and
$T_{\rm eff}\approx 19200$~K. We integrate the $I$-band luminosities
estimated from the extinction-corrected magnitudes. Assuming a constant
bolometric correction during the 16-day observing period, we obtain a
total (optical) radiated energy of $1.4-2.7\times 10^{49}$~erg and
luminosities of $1-2\times 10^{43}$~erg~s$^{-1}$ as of Oct.~9, depending
on the correction used. This is already comparable to the estimated
radiated energy of the Type~IIn SN~2011ht during its $\sim 100$~day
plateau phase (Roming et al. 2012; Mauerhan et al. 2012a). 

The evolution of SN~2009ip since the discovery of its latest 
recorded eruption on 2012 Jul.~24 (Drake et al. 2012) is quite
remarkable (Figure~3; see also M12b and P12), as it is unlike
any other Type~IIn SNe studied to date. The absolute magnitude was
approximately constant at $\sim -14.5$ ($V$ to $I$-bands) for a period of $\sim
50$~days after Jul.~24, and even faded to $\sim -13.5$ (at least between
Sep.~18 and Sep.~22.5, Martin et al. 2012; Margutti et al. 2012a) before
the rapid brightening that (given our observations) most likely started
between Sep.~23.7 and 24.5. If we assume the maximum bolometric
correction estimated above as an upper limit and a plateau phase at $M_I
= -14.5$ for $50$~days, we constrain the total radiated energy during
this period to be $\lesssim 5\times 10^{48}$~erg. However, we should note
that the brightening could have started earlier than Jul.~24, in which
case this could be a lower limit in the total radiated energy during
this phase.

The spectroscopic evolution of SN~2009ip since its latest recorded
eruption have also been unique. Foley et al. (2012) reported an optical
spectrum obtained on Aug.~26, which showed a blue continuum with narrow
($\sim 700$~km~s$^{-1}$) emission lines of H and He~I, similar to the
spectra obtained during the 2009 eruption (Smith et al. 2009; Foley et
al. 2012). Vinko et al. (2012) reported that an independent spectrum
obtained on Aug.~26 also showed broad P-Cygni features with very fast
velocities of $\sim 10,000$~km~s$^{-1}$, consistent with SN shock
speeds. These broad features seem to have persisted, and even
strengthened, until (at least) Sep.~23 (M12b). However, the high
velocity features and P-Cygni profiles had mostly gone away by
Sep.~26-27, leaving behind strong, narrow ($\sim 1,000$~km~s$^{-1}$)
emission features characteristic of many Type~IIn SNe spectra (M12b;
Burgasser et al. 2012; Vinko et al. 2012; Gall et al. 2012; P12).
 
As shown in Figure~3, the transition in spectroscopic properties is
coincident with the very rapid brightening observed in the optical light
curves. This, most likely, marks the start of a strong interaction
between the fast SN shock and a dense CSM environment likely formed from
material ejected in the previous LBV outbursts, as proposed by M12b. The
timing of the rapid optical brightening is also coincident with the
detection of X-ray emission from SN~2009ip (Campana \& Margutti 2012),
which is consistent with this picture. Another possible signature of a
strong ejecta-CSM interaction is the decline of $\sim 1$~mag observed
right before the spectral changes were seen and the light curve started
to rise rapidly. Moriya \& Maeda (2012) propose that this is an
unavoidable consequence of a fast SN shock breaking out of a dense CSM. 

Figure~4 shows a zoom-in view of the evolution of the optical luminosity
of SN~2009ip after Sep.~23.5, which is approximately when it started to
transition to a Type~IIn-dominated spectrum. For comparison, we also
show the optical luminosity evolution of three Type~IIn SNe from the
literature that have relatively well-sampled light curves before and
after peak brightness. This sample includes the very luminous SN~2003ma
(Rest et al. 2011) and SN~2006gy (Smith et al. 2007), and the relatively
low-luminosity SN~2011ht (Roming et al. 2012; Mauerhan et al. 2012a; but
see Humphreys et al. 2012). 

The very early rapid optical brightening of the Type~IIn SN~2009ip is
consistent with $L\propto t^2$ during a short 2-day phase, although this
naturally depends on the $t_{0}$ used. This is the evolution expected
for an homologously expanding (optically thick) black-body photosphere
at constant expansion velocity and effective temperature. The luminous
Type~IIn SN~2003ma and SN~2006gy show relatively similar evolution in
their optical luminosity at early times before peak. In fact, Smith \&
McCray (2007) modeled the early rise of SN~2006gy with c model.
SN~2011ht rises at a slower rate in the optical, although it was
observed to rise much faster in the UV-range (Roming et al. 2012). Given
the simplified assumptions that go into the  $L \propto t^2$ scaling, we
plan to explore more realistic models (e.g., Moriya et al. 2012) to fit
the early time light curve of SN~2009ip in a future publication. 

At later times, close to peak brightness, the optical luminosity and
light curve shape of SN~2009ip becomes more similar to SN~2011ht than
SN~2003ma or SN~2006gy. Mauerhan et al. (2012a) proposed that SN~2011ht
belongs to a growing sub-class of Type~IIn SNe (Type~IIn-P; including
also SN~2009kn, Kankare et al. 2012, and SN~1994W, Sollerman et al.
1998) with long-lasting post-peak plateaus of $\sim 100$~days ($M\sim
-18$~mag), and faint late-time decay slopes consistent with low
$^{56}$Ni yield production ($\sim 0.01$~M$_{\odot}$). Perhaps we are
seeing something similar in the case of SN~2009ip (but see P12),
although there are important differences in spectroscopic properties
compared to the small group of Type~IIn-P SNe (e.g., SN~2009ip shows
significantly hotter continuum and the narrow lines do not show P-Cygni
profiles). An interesting related issue that comes out from the recent
observations of SN~2009ip before the rapid brightening and fast
transition to narrow-line dominated spectrum, is whether surveys might
be missing a relatively low-luminosity phase (at least compared to the
main peak) in which the broad SN shock velocities are revealed. Given
the typical depth of transient surveys, this is certainly possible
(e.g., see Fig.~17 in Kiewe et al. 2012). In fact, Smith et al. (2011b)
proposed such an origin for the $\sim 100$~day plateau at $M\sim
-16$~mag observed before the main peak in the light curve of SN~1961V. 

\section{Conclusions}
\label{sec4}

We have presented high-cadence optical photometric monitoring of the
remarkable, young Type~IIn supernova SN~2009ip using data obtained
between 2012 Sep.~23 and Oct.~9 with $0.3$-$0.4$ meter aperture
telescopes from the Coral Towers Observatory (Cairns, Australia). Using
difference imaging analysis, we obtain precise $R$ and $I$-band light
curves and are able to resolve well-defined brightening phases. In
particular, we see a very rapid brightening early on (0.5~mag in 6~hr)
that quickly turns over after a couple of days. The SN brightened
between absolute magnitude $M_{I}=-14.2$ and $M_{I}=-17.9$ in 16~days.
The changes that we observe in the light curves are correlated with the
reported spectroscopic changes and the X-ray detection, and, most
likely, mark the start of a strong interaction between the fast SN
ejecta and the dense CSM formed from material that was ejected in the
pre-supernova LBV eruptions observed since 2009. Our study highlights
the importance of organized, high-cadence observations by amateur
astronomers with small-aperture telescopes and the impact they can have
in transients and SNe research, in this case with direct important
connections to massive stellar evolution.

\acknowledgments

We thank S.~de~Mink, O.~Graur, J.~Mauerhan, R.~Quimby, and A.~Rest for
stimulating discussions about SN~2009ip, and L.~Watson for detailed
comments on an earlier version of this manuscript. J.~L.~P. acknowledges
support from a Carnegie-Princeton Fellowship.

\newpage

\begin{figure}[t]
\plotone{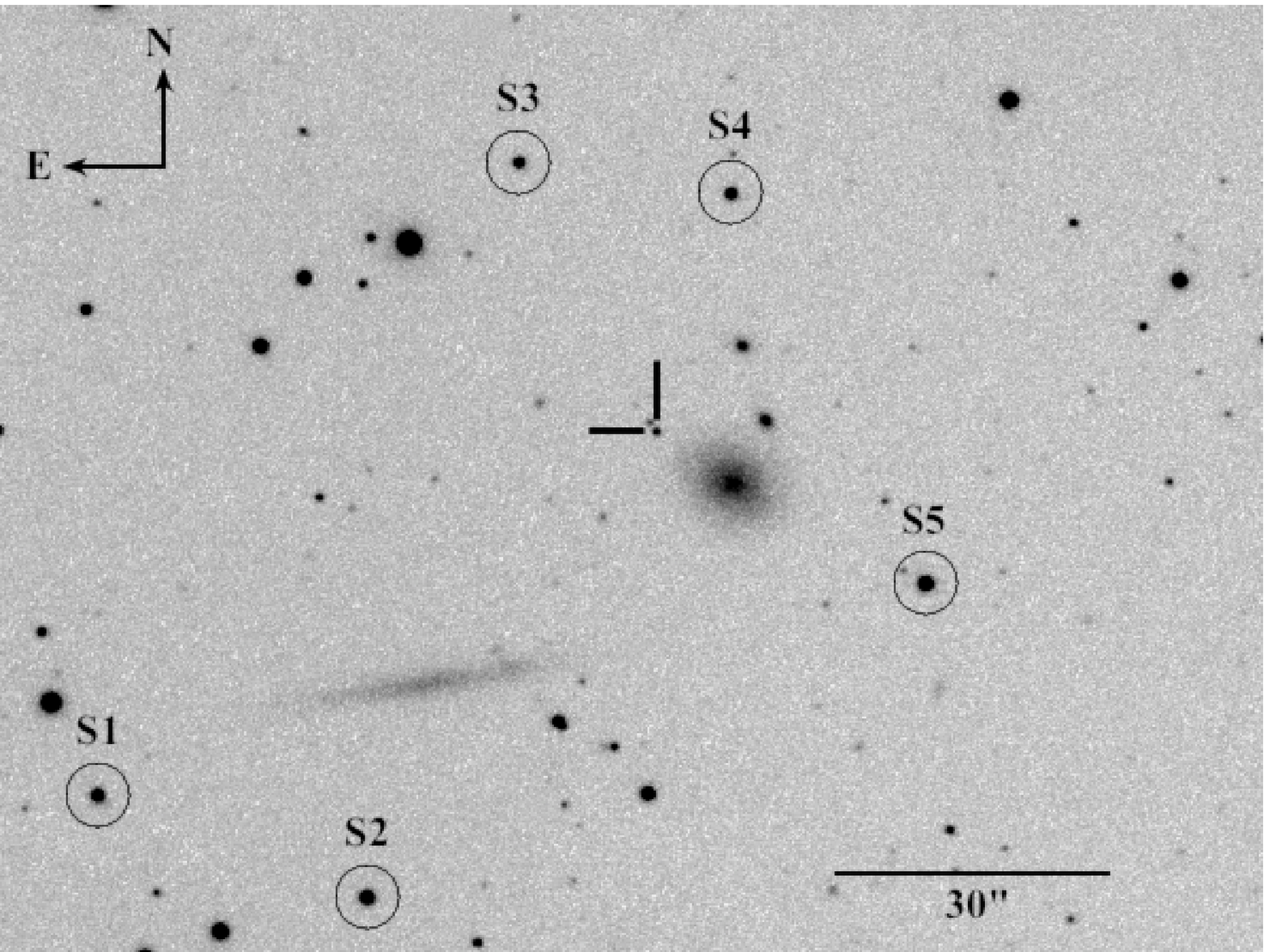}
\caption{Finding chart of SN~2009ip. This image was obtained with the IR filter on 
2012 Sep.~24. The circles mark the local standard stars used to estimate the photometric 
zeropoints and the cross-hairs show the position of SN~2009ip.}
\label{fig:fchart}
\end{figure}

\begin{figure}[t]
\plotone{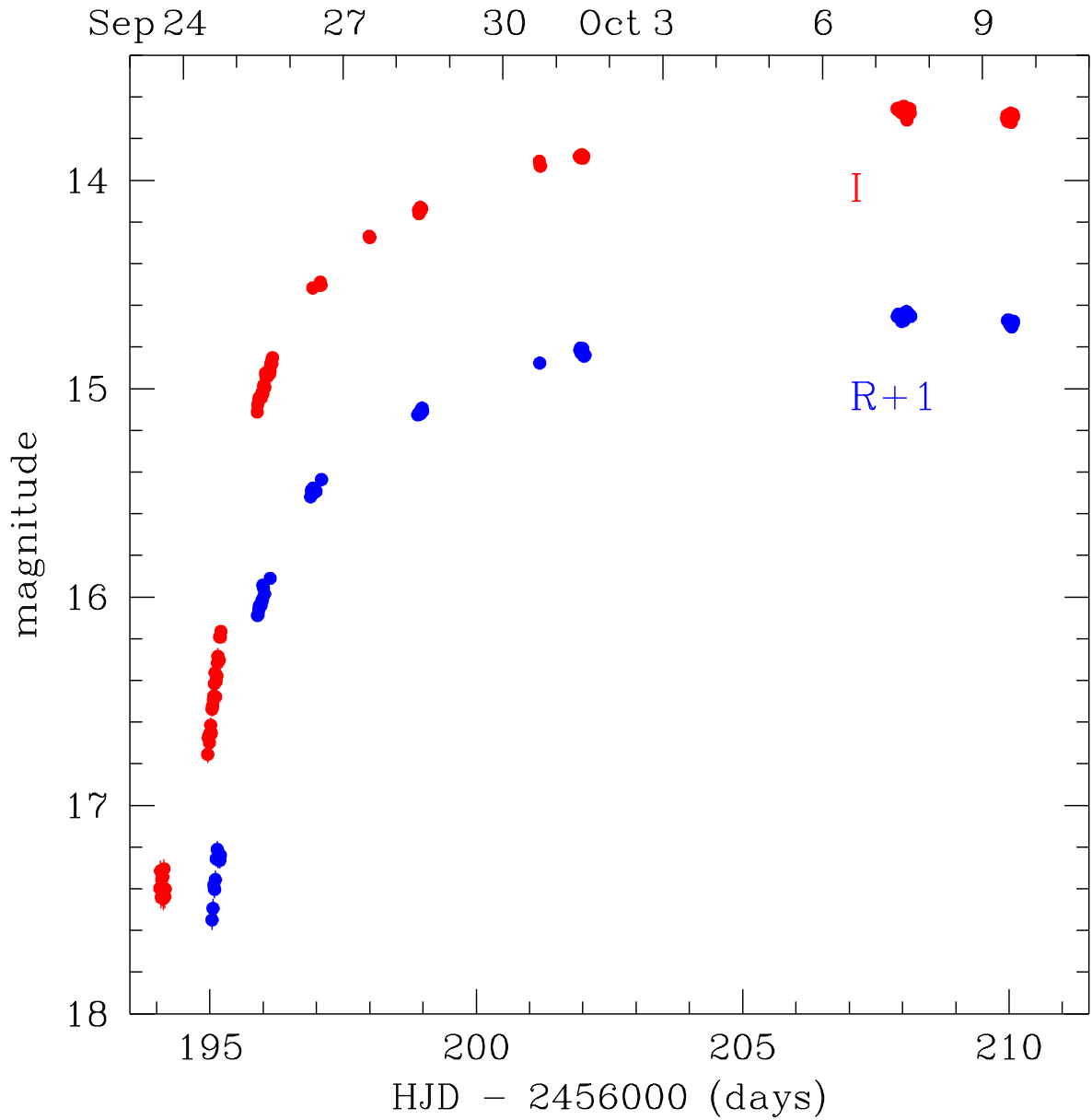}
\caption{Light curves of SN~2009ip in the $R$ and IR ($>
700$~nm) bands, calibrated to standard $R$ and $I$ magnitudes,
respectively. The data were obtained by J.~Brimacombe between 2012
Sep.~23.6 and Oct.~9.6 from the Coral Towers Observatory (Cairns,
Australia).}
\label{fig:lc}
\end{figure}

\begin{figure}[t]
\epsscale{0.9}
\plotone{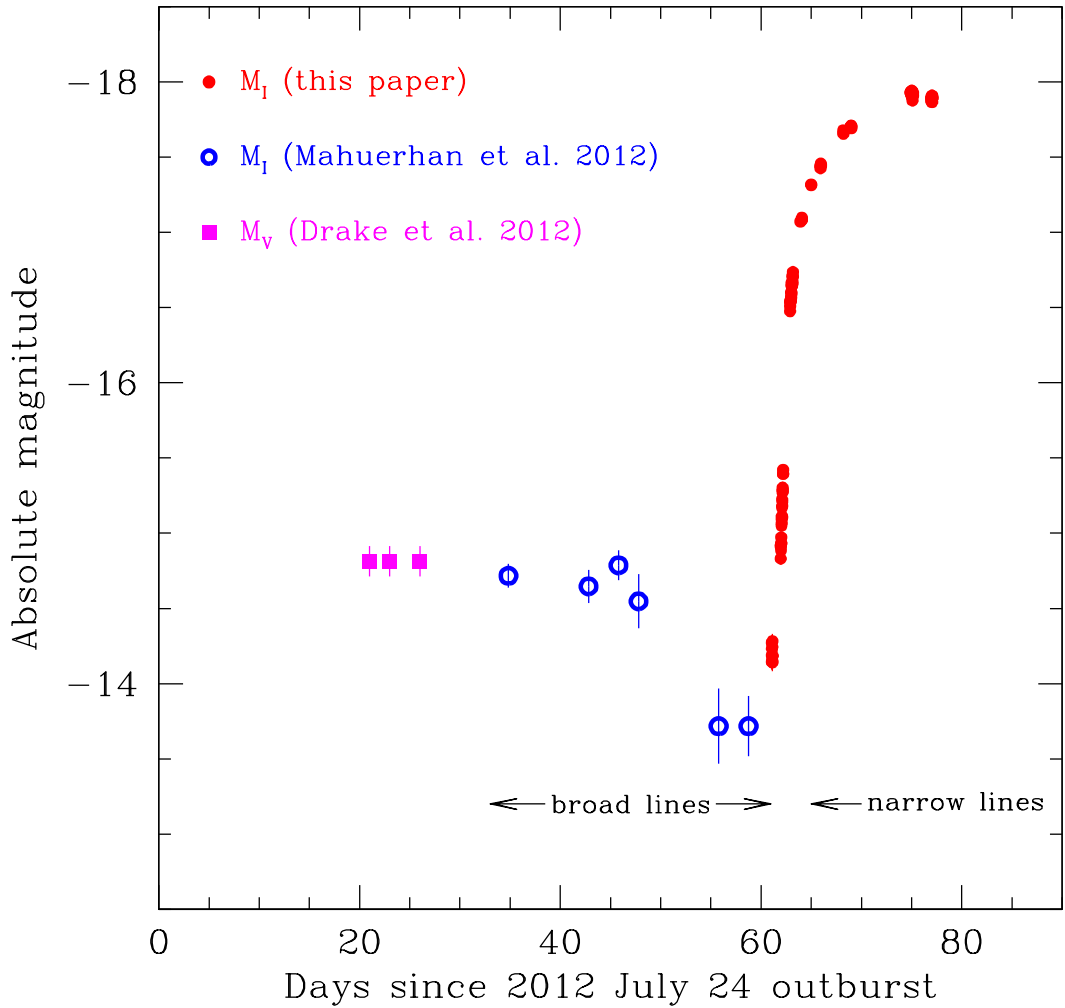}
\caption{Absolute magnitude light curve of SN~2009ip since its latest
eruption was reported on 2012 July~24 by Drake et al. (2012). The
filled circles are the $I$-band photometry presented in this paper; the
open circles are $I$-band photometry reported in Mauerhan et al.
(2012b); and the filled squares are $V$-band photometry reported in
Drake et al. (2012). The spectrum of SN~2009ip showed broad H and He~I
emission lines with P-Cygni profiles ($v\simeq 10,000$~km~s$^{-1}$),
characteristic of SN shock speeds, starting on (or before) Aug.~26
(Vinko et al. 2012; but see Foley et al. 2012) until, at least, Sep.~23
(Mauerhan et al. 2012). However, by Sep.~$26-27$ the spectrum had
changed substantially (Mahuerhan et al. 2012b; Burgasser et al. 2012;
Vinko et al. 2012) and the emission lines became strong and narrow
($v\sim 1,000$~km~s$^{-1}$), resembling the spectra of Type~IIn SNe.
These phases in the spectral evolution are shown schematically with
arrows. The change to a narrow-line dominated spectrum is coincident
with the very rapid brightening observed in the light curve.}
\label{fig:lcabs}
\end{figure}
\begin{figure}[t]

\plotone{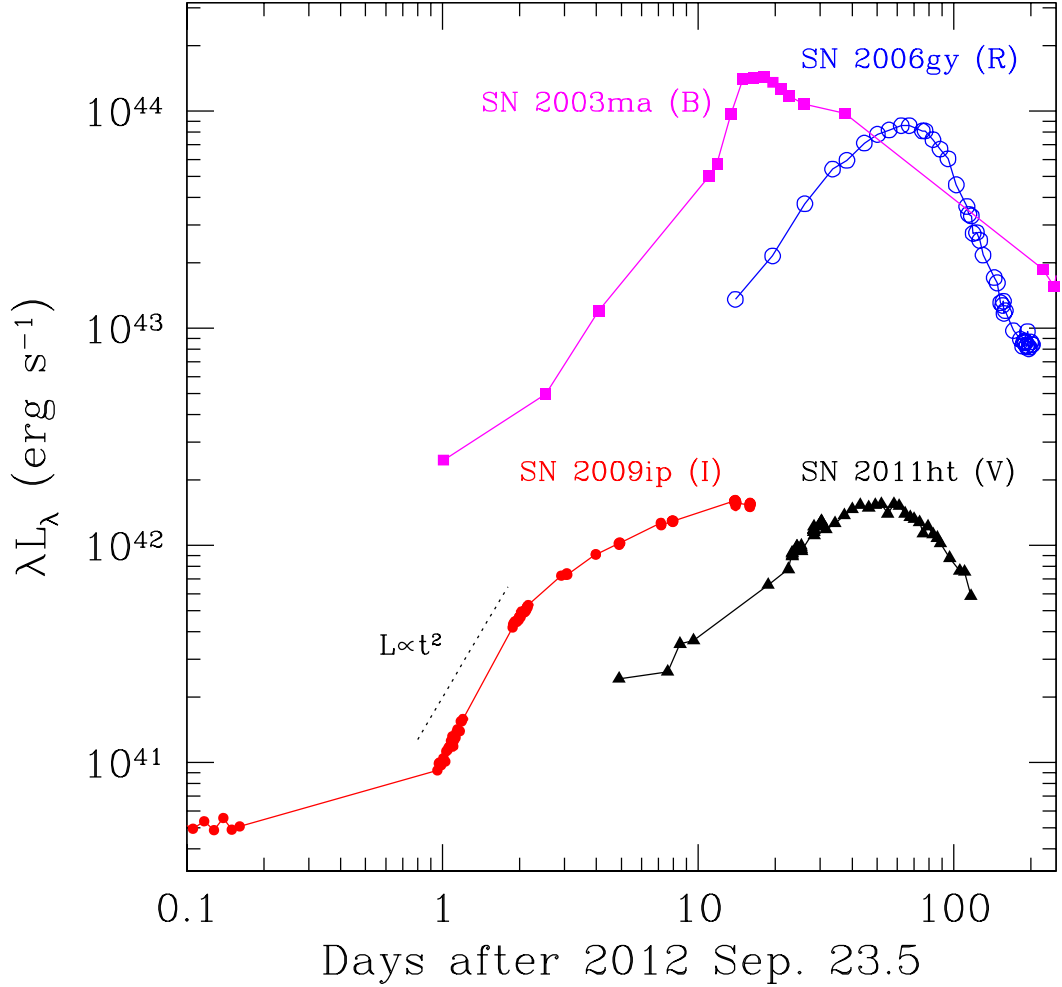}
\caption{Early rise in optical luminosity of the Type~IIn SN~2009ip
(filled circles) since 2012 Sep.~23.5. We also show for comparison the
luminosity evolution in the rest-frame optical of other Type~IIn SNe
(filters used indicated in parenthesis) with good light curve coverage
before and after peak brightness. These include: SN~2003ma (filled
squares; Rest et al. 2011), SN~2006gy (open circles; Smith et al. 2007),
and SN~2011ht (filled squares; Roming et al. 2012). We have used the
distances, extinctions, and $K$-corrections provided in each paper to
derive the luminosities. The time in these cases is estimated with
respect to an approximate explosion date (SN~2003ma and SN~2006gy) or
with respect to the discovery date (SN~2011ht). The dotted line shows
the luminosity scaling $L\propto t^2$.} 
\label{fig:lclum}
\end{figure}
\clearpage

\begin{deluxetable}{lcccc}
\tablecolumns{5}
\tablewidth{0pt}
\tablecaption{Light curve of SN~2009ip}
\tablehead{
\colhead{UT Date} &
\colhead{HJD - 2456000} &
\colhead{mag} & 
\colhead{$\sigma$} &  
\colhead{Filter}
}
\startdata
Sep 24.536400 & 195.040909 & 16.549 & 0.047 & $R$ \\ 
Sep 24.552130 & 195.056641 & 16.494 & 0.045 & $R$ \\ 
Sep 24.567801 & 195.072327 & 16.382 & 0.040 & $R$ \\ 
Sep 24.583472 & 195.087997 & 16.403 & 0.043 & $R$ \\ 
Sep 24.599144 & 195.103668 & 16.356 & 0.044 & $R$ \\ 
Sep 24.614826 & 195.119370 & 16.256 & 0.039 & $R$ \\ 
Sep 24.633634 & 195.138153 & 16.211 & 0.040 & $R$ \\ 
Sep 24.644965 & 195.149475 & 16.259 & 0.043 & $R$ \\ 
Sep 24.656285 & 195.160797 & 16.256 & 0.040 & $R$ \\ 
Sep 24.667593 & 195.172134 & 16.250 & 0.038 & $R$ \\ 
\ldots & \ldots & \ldots & \ldots & $R$ \\ 
\enddata
\tablecomments{The full table will be available as a machine readable table.}
\end{deluxetable}

\end{document}